\pdfoutput=1 
\documentclass[12pt]{article}
\usepackage{amsmath,amsfonts,amssymb,a4,epsfig,color,graphics}

\newcommand{\R}{{\mathbb{R}}}

\newcommand{\N}{{\mathbb{N}}}

\def\ha{\frac{1}{2}}

\def\ra{\rightarrow}
\def\preuve{\begin{proof}}

\def\gd{\delta}
\def\ge{\varepsilon}

\def\gg{\gamma}

\def\gl{\lambda}

\newtheorem{defi}{Definition}

\newtheorem{lemm}{Lemma}

\newtheorem{rem}{Remark}
\newtheorem{coro}{Corollary}
\newtheorem{theo}{Theorem}
\newtheorem{exem}{Example}[section]
\newenvironment{demo}{\noindent {\it Proof.--}
      \begin{quotation}\noindent}{\end{quotation}\hfill$\square $}

\begin{document}

\title{Tunneling on graphs:\\
an approach ``\`a la Helffer-Sj\"ostrand''}
\author{Yves Colin de Verdi\`ere \footnote{Institut Fourier,
 Unit{\'e} mixte
 de recherche CNRS-UJF 5582,
 BP 74, 38402-Saint Martin d'H\`eres Cedex (France);
yves.colin-de-verdiere@univ-grenoble-alpes.fr}
}


\maketitle



\section*{Introduction}
In the paper \cite{LLY}, the authors study the tunneling
effect on a finite graph $G$. In order to evaluate the eigenvalues
of a  Schr\"odinger operator on $G$ 
 in the semi-classical regime, they introduce
a kind of Dirichlet to Neumann map which gives an implicit
equation. 
On the other hand, Bernard Helffer and Johannes Sj\"ostrand gave a very 
 explicit 
approach to the estimation of the eigenvalues
of a semi-classical Schr\"odinger operator in $\R^d$ in several papers.
In particular, in \cite{HS1}, they introduce the so-called 
interaction matrix whose eigenvalues are close to the tunneling
eigenvalues.

The goal of this note is to show that the Helffer-Sj\"ostrand approach 
is also suitable for the problem on graphs and to describe how
to compute explicitely the interaction matrix. 

\section{The problem}

We consider  a finite non-oriented graph $G=(X,E)$ with no loops and 
 we denote by $d(x)$ the degree of the vertex $x$
and by  $D(x,y)$ the combinatorial distance
between the vertices  $x$ and $y$.   A  Schr\"odinger
operator $H$  on $G$ is defined by 
\[ H=\hbar^2 \Delta + V ~,\]
where 
\begin{itemize}
\item 
$ \hbar $ is a positive parameter. The {\it semi-classical limit} that
we will study is $\hbar \ra 0$.  
\item $\Delta $ is the linear symmetric operator
on $\R^X$ defined
by \[\Delta f(x)=-\sum_{y\sim x}f(y))~.\]
\item The potential $V$ is a function $V:X\ra [0, \infty[ $.
$V$ is called {\it simple} if $\forall x\in X, ~ V(x)\in \{ 0,1 \}$.
\item A {\it well} $x\in X$ is a vertex of $G$ so that $V(x)=0$.
$L=\{ 1,\cdots,j,\cdots,N \}$ denotes the set of wells.
We assume in what follows that there is no edges between 2 wells.
It means that the wells are isolated vertices of $G$.
\end{itemize}

\section{Dirichlet problems and decay estimates}

Let $j \in L$ and $L_j=L\setminus \{ j \} $.
We will consider the restriction $H_{j}$
of $H$ to the space of functions $f:X\ra \R$ which vanish
on  $L_{j}$.
The ground state of $H_j$ is   a function $\psi_j $
which is $>0$ on $X\setminus L_j$. We normalize $\psi _j$
by $\psi_j(j)=1$.  The associated eigenvalue is denoted
$\mu_j$. 
\begin{lemm}
As $\hbar \ra 0$, $\lim _{\hbar \ra 0} \psi_j $
is the function $\ge_j$ defined by $\ge_j(j)=1$
and $\ge_j (x)=0$ is $x\ne j$
and $\lim_{\hbar \ra 0} \mu_j =0 $.
Moreover $\psi _j$ and $\mu_j$ are analytic functions of $\hbar^2$.
\end{lemm}
This is clear because the matrix of $H_j$ is analytic in $\hbar ^2$
and the limit for $\hbar=0$ is a diagonal matrix with all entries
$>0$ except the $j$-th which is $0$.

{\it Some notations:}
if $P=(x_0,\cdots , x_{|P|})$ is a path, we define the weight $s_\gl(P)$
by
\[ s_\gl (P)=\hbar^{2|P|}a(x_0)\cdots a(x_{|P|-1})\]
with 
\[ a(x)=(V(x)-\lambda )^{-1}~.\]
Let us note that $s_\gl(P)$ depends  on $\hbar$.
Sometimes it will be convenient to write
$s(P)=s_\lambda (P)$. 

\begin{theo}\label{prop:psij}
Let us define, for $\lambda $ close to $0$,
 the function $\psi_\gl $ by $\psi_\gl(j)=1$, $(\psi_\gl) _{|L_j}=0$
 and, for $x\notin L$,
\[ \psi_\gl  (x)=\sum _{P:x \ra j} s_{\lambda}(P)~,\]
where the (convergent) sum is on all paths with $x_0=x$, $x_{|P|}=j$ and
$x_l \notin L$ for $1\leq l \leq |P|-1$, 
Then 
$\mu_j$ is defined implicitely by
\[ \sum _{P:j \ra j}s_{\mu_j}(P)=1~,\]
where the sum is on all paths with $x_0=j$, $x_{|P|}=j$ and
$x_l \notin L$ for $1\leq l \leq |P|-1$.

In particular $\mu_j=\hbar^4 \sum _{y\sim j}V(y)^{-1} +O\left(\hbar^6\right)$,
$\| \psi_j \| =1 +O\left( \hbar^4 \right)$
and
\[ \forall x\in X,~\psi_j(x)=O\left(\hbar^{2D(x,j)}\right)~.\] 
\end{theo}
\begin{rem}
The implicit equation for $\mu_j$ can be expanded
as
\[ \mu_j = -\sum_{k=2}^\infty  \hbar^{2k} \sum _{P=(j,x_1,
  \cdots, j),~|P|=k }
\prod_{l=1}^{|P|-1} (V(x_l)-\mu_j )^{-1}~.\]
This equation can be solved induction.
This is related to the so-called Rayleigh-Schr\"odinger series.
\end{rem}
\begin{demo}The sums on paths are absolutely convergent for $\hbar $
small enough because of the following upper bound:
\begin{lemm} If $G=(X,E)$ is a finite graph and $x\in X$, the
  number
of paths of length $l$ starting from $x$  is bounded from above 
by $\left(\max_{x\in X} d(x)\right)^l$. 
\end{lemm}
So, for $\gl $ close to $0$ the series defining $\psi $
is bounded by $O\left(\sum _l (C\hbar)^{2l} \right)$.

Let us show first that $\psi_\gl $ satisfies
$((H-\lambda )\psi_\gl )(x)=0$ if $x\notin L$.
We have
\[ ((H-\lambda
)\psi_\gl)(x)=\frac{1}{a(x)}\psi_\gl(x)-
\hbar^2\sum_{y\sim x,~y\notin L_j}\psi_\gl(y)~,\]
and, using the definition of $\psi_\gl$, the last sum
is 
$\sum _{y\sim x,~y\notin L_j }\sum_{Q:y\ra j }s_\gl(Q)$.
Using the decomposition of $P:x\ra j$ as a path 
$(x,Q)$, we get
\[ ((H-\lambda
)\psi_\gl)(x)= 0~.\]

Similarly we can compute $((H-\lambda
)\psi_\gl)(j)$ as
\[((H-\lambda)\psi_\gl)(j)=\frac{1}{a(j)} -
\hbar^2\sum _{y\sim x}\sum_{Q:y\ra j }s_\gl(Q)=
\frac{1}{a(j)}\left(1 -\sum _{P:j\ra j}s_\gl (P)\right)~.\]

\end{demo}

\section{The interaction matrix}
Our goal is to apply Theorem \ref{theo:interaction}
with ${\cal F}$ the space generated by the $\psi_j$'s with $j\in L$. 
Using Proposition \ref{prop:psij}, we can take
$\eta=\hbar^4$ and $\ge=\hbar^{2S_0}$ with
 $S_0:=\min_{i,j\in L,~i\ne j} D(i,j)$.
The diagonal entries of the interaction matrix 
$H_{\cal E}$ are the $\mu_j $'s estimated in Proposition
\ref{prop:psij}.
We need to compute
$\langle r_j | \psi_i \rangle $. 
Using the fact that $((H-\mu_j)\psi_j)(x)=0$ if $x\notin L\setminus
j$, we get
$\langle r_j | \psi_i \rangle=\sum _{l\in L\setminus
  j}((H-\mu_j)\psi_j)(l)
\psi_i(l)$.
We have
$((H-\mu_j)\psi_j)(l)=\sum_{P:l\ra j}\tilde{s}_{\mu_j}(P)$
with 
$\tilde{s}_{\mu_j}((l,x_1,\cdots, j))=-\hbar^{2|P|}\prod_{l=1}^{|P|-1}a(x_l)$.
We get
\[ \langle r_j | \psi_i \rangle=\sum _{P:i\ra j,~|P|=S_0}\tilde{s}_{\mu_j}(P)+
O\left(\hbar^{2S_0+2}\right)~.\] 
Summarizing, we get the
\begin{theo}
Up to $O\left( \hbar^{2S_0+ 2} \right)$, the $|L|$ first
eigenvalues
of $H$ are those of the matrix
$I = {\rm Diag}(\mu_j )+ r_{ij} $ 
with
\[ r_{ij}=-\hbar^{2S_0} \sum _{P=(i,x_1,\cdots , x_l,\cdots, j),~ |P|=S_0}
\prod _{l=1}^{|P|-1}\frac{1}{V(x_l)}
~,\]
where the paths $P$ in the sum satisfy $x_l \notin L$ for 
$1\leq l \leq |P|-1 $.
\end{theo}

\section{  Simple potentials on graphs of constant degree $d$}

\begin{defi} The potential $V$ is called {\rm simple}
if, for all vertices $x\in X$, we have $ V(x)=0$ or $1$.
\end{defi}
If we assume moreover that the vertices of $G$ are all of the 
same degree $d$, the matrix $I $ becomes purely combinatorial.

In this case, we have
\[ s(P)=\hbar^{2|P|}(  -\lambda )^{-1}
 (1 -\lambda )^{1-|P|}~,\]
and the equation for $\mu_j$ is
\[\mu_j=- \sum _{k=2}^\infty \hbar^{2k} (1 -\mu_j
)^{1-|k|}N_j(k)\]
where $N_j(k)$ is the number of paths $P:j\ra j $ of length $k$.


The non-diagonal entries of $I$ are given
by
 \[ r_{ij}=- \hbar^{2S_0}\# \{ P:i\ra j |~|P|=S_0  \}~.\] 

\section{Application to simulated annealing}

The problem is to find the global minimum of a function $H$
on a finite, but large set $X$.
We assume that the set $X$ has a graph structure $G=(X,E)$ which gives a way
to move on it.
\begin{exem}$X$ is the set of element of the group ${\cal S}_N$ of 
permutations of $N$ letters. $S$ is a small generating set of ${\cal
  S}_N$
and the $G$ is the associated Cayley graph.
\end{exem}
\begin{exem}$X=\{-1,+1\}^Y $ is a spin system on the lattice $Y$ and, if
$x,y \in X$,  
$\{x,y\}\in E$ if all cordinates of $x$ and $y$ are the same except one. 
\end{exem}
 The function $H$ can be assumed to be with values
in $\N $ and we can also assume that, for $\{ x, y\} \in E $,
$H(x)-H(y)=\pm 1 $.
Let us fix some positive number $T >0$, the temperature, then
there is a  probability measure  on $X$,
 called the Gibbs measure, defined
by
$\mu _T (\{x \} )=Z^{-1}e^{-H(x)/T}$.
As $T \ra 0^+ $, the measure $\mu_T$ is more and more 
concentrated on the global minima of $H$.
We can define a Markov process   on $X$ 
by the transition matrix $\Lambda _T $ defined
by 
 $\gl_{x,y}=1$ if $H(y)< H(x)$,   $\gl_{x,y}=e^{-(H(y)-H(x))/T}$
if $H(y)>H(x)$ and  $\gl_{x,x}=-\sum_{y\sim x}\gl_{x,y}$.
 The quadratic form associated to $-\Lambda _T$ is 
\[ q_T(f)=\ha Z^{-1}\sum_{x\in X} e^{-H(x)/T}\sum _{y \sim x}
\gl_{x,y}(f(x)-f(y))^2 ~.\]
The measure $\mu_T$ is the stationary measure of this Markov process
defined 
by 
\[ {\rm Prob}( \{ \gg |  \gg (0)=x, \gamma (t) =y \})=
e^{t\Lambda _T^\star }(x,y)~.\]

The matrix $\Lambda _T $ gives a symmetric map on $l^2(\mu _T)$ whose
eigenvalues
are $\gl_1 =0 > \gl_2 \geq \cdots $ .
The speed of convergence of a random trajectory is basically
controlled by the gap $-\gl_2$  of the matrix $\Lambda _T $.
The main information is given by the asymptotic behavior of the gap
as $T\ra 0^+$. This asymptotic behaviour is the main object of the
paper \cite{CPY}.
In this paper, we propose an algorithm in order to determine
the order of magnitude of the gap: an even power of
$\ge=e^{-1/T}$.

The first step is to indentify
$l^2(X,\mu_T)$ with $l^2(X,{\rm can}) $ where ${\rm can}$ is the
measure $\sum _{x\in X} \gd (x) $.
This is done using the unitary map
$U:l^2(X,{\rm can})\ra l^2(X,\mu_T)$
defined by 
\[ (Uf)(x)=Z^{\ha}e^{H(x)/2T}f(x) ~.\]
The quadratic form associated to $H_T=-U^\star \Lambda _T U $,
$ Q_T(f)= q_T (Uf)$ is given by
\[ Q_T(f)=\sum _{  x\sim y, ~H(y)=H(x)-1}\left( f(x)- \ge f(y)
\right)^2 \]
with $\ge ={\rm exp}(-1/2T )$.
It can be checked that the lowest eigenvalue of $Q_T$ is $0$ with
eigenvector $f(x)=\ge ^{H(x)}$ which concentrate on the global minimas
of $H$.
We have also
$H_T= -\ge A_G + V_\ge $
where $A_G$ is the adjacency matrix 
and $V_\ge (x)=n_+ (x) +\ge^2 n_- (x)$ with
$n_+(x)=\# \{ y\sim x|H(y)=H(x)-1 \}$
and $n_-(x)=\# \{ y\sim x|H(y)=H(x)+1 \}$.

Our goal in \cite{CPY} was do determine the asymptotic behavior of the gap of $H_T$
as $T \ra 0^+$. This can also be done using the previous approach with $\hbar :=\sqrt{\epsilon}$ and $V$ depending now of $\hbar$
in a smooth way.

\section*{Appendix A: abstract interaction matrix}

Let  ${\cal H}$ be an  Hilbert space (assumed to be real for
simplicity)
 and  ${\cal E},{\cal F}$
two  subspaces of ${\cal H}$, let us define the  ``distance'' 
\[ d({\cal E},{\cal F})=
\sup _{x\in {\cal E},~\|x\|=1} \inf _{y\in {\cal F}} \| x-y \|~.\]
If $\dim {\cal E}=\dim {\cal F}=N<\infty$, one checks, using an
 isometry of ${\cal H}$ exchanging
${\cal E}$ and ${\cal F}$, that $d$ is symmetric.

\begin{lemm}\label{lemm:dist}
Let $A$ be self-adjoint on ${\cal H}$,
$I=\lbrack \alpha , \beta  \rbrack \subset \R $
and  $a>0$ so that  ${\rm Spectrum(A)}\cap (\lbrack \alpha -a, \alpha  \lbrack
 \cup \rbrack \beta , \beta +a \rbrack )=\emptyset   $.
Let  $\psi _j,~j=1,...,N, $ so that 
\begin{equation}\label{equ:quasimodes}
\| (A-\mu _j)\psi _j \| \leq \varepsilon 
\end{equation}
 with $\alpha \leq \mu _j
\leq \beta  $ and  ${\cal F}$ the space generated by the  $\psi _j $'s.
If  ${\cal E}$ is the range of the
spectral projector  $\Pi $ of  $A$ associated to the interval
$I=\lbrack \alpha ,\beta \rbrack $, we have:

$$ d({\cal F},{\cal E})\leq {\varepsilon \sqrt{N} / a\sqrt \lambda _S }~,$$
where  $\lambda _S$ is the smallest eigenvalue of the matrix
$S=(s_{ij})=(<\psi _i \vert \psi _j >)$.
\end{lemm}

\begin{demo}
Let  $\psi _j =v _j +w _j$ where  $v _j$ is the  projection of
$\psi _j $ on ${\cal E}$.
We have, using the fact that  $w_j$ belongs to the image of
 the spectral projector
${\rm Id}-\Pi $ and  the assumption on the spectrum of $A$, 
$$\varepsilon \geq \| (A-\mu _j)\psi _j \| \geq \| (A-\mu _j)w _j \|
\geq a \| w _j \| $$
and hence  $\| \psi _j - v _j \| \leq {\varepsilon  /a}$.

If  $\psi =\sum x_j \psi _j $ and  $v=\sum x_j v _j  $ is the  projection of 
$\psi$ on ${\cal E}$,
we have, using Cauchy-Schwarz inequality, 
$$ \| \psi - v \| \leq \sqrt{\sum x_j ^2}.\sqrt{N}.{\varepsilon \over a}~, $$
and:
$$ \| \psi \| ^2 = \sum x_ix_j s_{i,j} \geq \lambda _S \sum x_j ^2
~.$$
The result follows.
\end{demo}

We keep the Assumptions of Lemma
\ref{lemm:dist}, in particular Equation (\ref{equ:quasimodes}),
 and  assume now that 
  $\dim({\cal E})=\dim({\cal F})=N$ so that $d({\cal E},{\cal F})=0(\epsilon)$.
 We assume also that  we have two small parameters
$ \eta =o(1),~\epsilon=o(1)$ and that 
\begin{equation}\label{equ:psiscal}
\langle \psi_i | \psi_i \rangle =
1 + O(\eta) {\rm ~and~ for~} i\ne j,~ 
\langle \psi_i | \psi_j \rangle =O(\epsilon)~.
\end{equation}
 We denote by
$\Psi_i=\psi_i/\| \psi_i \|$,
$V_i=\Pi \Psi_i$. If  $\Sigma $ is the matrix of the scalar
products   $\Sigma=(\langle V_i |V_j
 \rangle )$ and if $(\kappa_{ij})$ denotes the  matrix
$\Sigma^{-1/2}$, we put
$e_i=\sum_k \kappa _{ik}V_k$. The set ${\cal O}=\{ e_i|i=1,\cdots, N\}$
is an orthonormal basis of ${\cal E}$.
The next statement gives an approximation of the matrix of
the restriction $A_{\cal E} $ of  $A$ to ${\cal E}$
in the basis ${\cal O}$:

\begin{theo}\label{theo:interaction}
 The matrix  $A_{\cal E} $ of  $A_{|{\cal E}}$ in the basis  ${\cal O}$
 is given by:
$$a_{ij}= <Ae_i \vert e_j > = \mu _i \delta _{i,j}
+{1\over 2}\left(<r_i \vert \psi _j >+<r_j \vert \psi _i >\right)
+O\left(\varepsilon (\ge+ \eta) \right)~, $$
with  $r_i =\left(A-\mu _i\right)\psi _i =O\left(\varepsilon \right)$.
\end{theo}

\begin{demo}

First, by  Pythagore's Theorem and using Equation (\ref{equ:psiscal}), 
$$ <V_i \vert V_j >= <\Psi _i \vert \Psi _j > +O\left(\varepsilon ^2 \right):=
 \delta _{i,j}+T_{i,j}$$
with $T=(T_{ij})=0(\epsilon)$.

Similarly
$$ < AV_i \vert V_j > = <A\Psi _i \vert \Psi _j > +
 O \left(\varepsilon  ^2 \right)~:$$
we start with $\Psi_i =V_i +W_i$ and 
$A\Psi_i =AV_i + AW_i$.
Using $A\Psi_i =\mu_i \Psi_i +r_i/\|\psi_i \| $
 and projecting on ${\cal E}^\perp $,
we get $AW_i =O(\ge)$.

We get then 
using the symmetry of $A$:
$$ \left(<AV_i \vert V_j >\right)= D_\mu 
 +{1\over 2}\left(D_\mu T+T D_\mu \right)+
{1\over 2}\left(<r_i \vert \psi _j > +<r_j \vert \psi _i >\right)
+O\left(\varepsilon (\ge+ \eta ) \right)~,$$
where  $D_\mu $ is the diagonal matrix
whose entries are the  $\mu _i $'s.

Using the fact that
$(e_i) =\left(Id - T/2 +O\left(\varepsilon ^2\right)\right)(V_j) $,
we get :
$$\left(<Ae_i \vert e_j >\right)=\left(Id -
{T/ 2}\right)\left(<AV_i \vert V_j >\right)\left(Id - {T/ 2}\right)
+O\left(\varepsilon ^2 \right)~. $$
The final result follows.\end{demo}

\begin{coro}
If $\gl_1 \leq \cdots \leq \gl_N $ are the eigenvalues of $A$
in the interval $I$ and  $\mu_1 \leq \cdots \leq \mu_N $
are the eigenvalues of
\[ D_\mu +\ha \left( \langle r_i|\psi_j \rangle
+\langle r_j|\psi_i \rangle \right)~,\]
then
\[ \gl_j =\mu_j +O\left(\ge (\ge+ \eta) \right)~.\]
\end{coro}

\bibliographystyle{plain}

\end{document}